# Use of high throughput sequencing to observe genome dynamics at a single cell level.

Parkhomchuk D[1], Amstislavskiy VS[1], Soldatov A[1] and Ogryzko V[2]

1: Department of Vertebrate Genomics, Max Planck Institute for Molecular Genetics, Berlin, Germany

2: CNRS UMR 8126, Université Paris-Sud 11, Institut Gustave Roussy, Villejuif, France

**Summary**

With the development of high throughput sequencing technology, it becomes possible to directly analyze mutation distribution in a genome-wide fashion, dissociating mutation rate measurements from the traditional underlying assumptions. Here, we sequenced several genomes of *Escherichia coli* from colonies obtained after chemical mutagenesis and observed a strikingly nonrandom distribution of the induced mutations. These include long stretches of exclusively G to A or C to T transitions along the genome and orders of magnitude intra- and inter-genomic differences in mutation density. Whereas most of these observations can be explained by the known features of enzymatic processes, the others could reflect stochasticity in the molecular processes at the single-cell level. Our results demonstrate how analysis of the molecular records left in the genomes of the descendants of an individual mutagenized cell allows for genome-scale observations of fixation and segregation of mutations, as well as recombination events, in the single genome of their progenitor.

**Introduction**

Cells can copy their genetic material with exceptional accuracy (the spontaneous mutation frequency in *E. coli* being as low as $4*10^{-10}$ base substitution mutations per base pair (bp) per generation). The robust amplification of the effects of an individual molecular event resulting from such accuracy has long set genetics apart from biochemistry as a discipline able to study individual events (such as mutations or recombinations) at the level of a single organism. Yet, until recently, the studies of genetic variability in living cells had been limited to very few genetic systems and typically relied on various selection screens (1). Making genome-scale inferences from these experiments requires the assumption of uniform event distribution, which is highly questionable due to the phenomenon of mutation hot spots (1,2). Also, in light of the phenomenon of adaptive mutations (3-5), it is preferable to study genetic variability with methods that do not depend on environmental context, as the emergence of mutations can be influenced by the selection conditions, in a still poorly understood fashion.

In addition, both mutation and recombination events could arise from the same set of circumstances. For example, a mutagen often also induces DNA lesions that obstruct DNA synthesis and cause collapse of the replication fork, which has to be repaired by homologous recombination (6). However, to study both of these effects of mutagenic treatment with conventional genetic methods in a single experiment is complicated.

Recent advances in high throughput genomic analysis open up new opportunities for analysis of genome variability (7-10). In particular, by expanding mutation analysis to the genome-wide scale, modern high-throughput sequencing technology permits us to detect correlations between individual molecular events in a single organism, independent of enhancement schemes. In our work, we set out to explore how this analysis can help in observing, at the single-cell level, the contributions and interactions between the molecular processes contributing to mutation generation and segregation.

**Results**

Ethyl methanesulfonate (EMS) was chosen as an efficient mutagen with excellent preservation of viability (11). The K-12 CC102 strain was chosen as a widely used model for mutagenesis studies for consistency of analysis (12). Bacteria were mutagenized according to a standard protocol (12) but, in order to minimize the loss of slightly deleterious mutations, the cells were grown for only 2 hours in rich medium before plating on LB agar. The next day, several colonies were picked at random and grown in LB for two more hours to obtain amounts of DNA sufficient for high-throughput sequencing analysis.

Illumina GA sequencing uses four fluorescently-labeled modified nucleotides to sequence by synthesis the tens of millions of clonal clusters, generated by fragmentation of DNA and amplification of the fragments via ligation-mediated PCR (see Materials and Methods) (13). First we describe general features of the data obtained from the sequencing of six clones of mutagenized CC102 cells. Consistent with the fact that the cells were proliferating at the moment of harvest, we observed a gradient in the sequencing coverage (Figure 1A) for every DNA sample, with the replication terminus (Ter) noticeably underrepresented compared to the replication origin (OriC). Another global genomic feature can also be inferred from this simple analysis of coverage: the region including the *proAB* and *lacZ* loci shows strong variability in coverage between different sequenced colonies (Figure 1A shows the coverage for one of the colonies), reflecting its independent replication as a part of the F´128 episome (14).

The single nucleotide differences between the parental CC102 strain and the K12 MG1655 reference strain are described below (see Figure 1B for an example of mutation identification, and the supplementary table ST1, left column, for counts). As expected, the following mutations were found, consistent with the genetic background of the CC102 strain: stop codon in the *araC* gene, *lacZ* GAG to GTG (Glu to Gly) in the 461 position of the *lacZ* gene and a promoter mutation in the *lacI* gene. All possible nucleotide replacements are present, reflecting the complicated laboratory histories of these strains after their divergence.

Sequence comparisons between the un-mutagenized parental CC102 strain and six descendants from the individual colonies after EMS mutagenesis revealed a picture dramatically dissimilar to the MG1655 versus CC102 comparison (table ST1, central and right columns). On average, 70 mutations per genome were observed. The overwhelming majority of the changes are G:C → A:T transitions, well in accord with the known mutagenic specificity of EMS (15-19). The genome-wide distribution of the newly acquired mutations induced by EMS in the CC102 strain is strikingly nonrandom and exhibits several prominent features, present in all six DNA samples that were sequenced.

The first feature is the presence of long stretches of genome where either only G → A or only C → T transitions are observed (Fig.2A, B, S1). The asymmetric stretches span up to 2 Mb, and often are switched to a stretch of the opposite kind (on average 8-10 times per genome sequenced, including very short switches within a stretch). The positions of the asymmetric stretches and how they switch vary between the six DNA samples sequenced, and do not manifest any regularity or correlation with a known genomic feature. We refer to these features as 'asymmetric stretches' and 'switches', respectively.

The second feature is a number of positions having a mixture of a mutated and wild type sequences in different reads (an example of such a position is shown on Figure 1B (bottom), and the genome-wide distribution on Figure 2B and S1). A particularly high number of these positions was observed for the colony H1. Two arguments suggest that this observation is not an artifact of the sequencing methodology: the sequencing coverage is high, and no such mixtures are detected when the MG1655 and the parental strain CC102 strain are compared. We will call this feature 'mixed states'.

The third feature of the mutation distribution is a striking difference between mutation densities in different regions of the genome. In fact, we detected two separate aspects of the uneven distribution. The regions of the genome positioned at two axes – the Ori-Ter axis and its orthogonal axis – consistently displayed lower mutation density (with up to an order of magnitude difference) in all colonies sequenced (Figure 2C). However, we also detected regions (examples are indicated in Figure 2B) that showed dramatic and statistically significant individual variations from one sequenced colony to another (Figure 2D). We term this aspect of uneven mutation distribution 'mutation bunching'.

How might these striking inter-genome and intra-genome irregularities in mutation distribution be explained? Albeit somewhat surprising, and strongly deviating from the unbiased genome-scale distribution naively expected from a mutagenic process, asymmetric stretches and switching can be accounted for by the following 'fixation and segregation' model based on the semiconservative model of DNA replication and knowledge of how DNA lesions are converted into mutations or induce recombination events.

1. The asymmetric stretches of G → A or C → T transitions are straightforward to explain. Figure 3A (top) shows the standard model of the EMS-induced $O^6$ alkyl guanine specifically mispairing with thymine, resulting in a G to A replacement upon the second round of replication (15-19). Considering now a continuous stretch of DNA, and assuming that each strand in the original DNA is randomly affected by EMS, one would expect that the segregation between daughter strands into different cells after replication will lead to each descendant cell having exclusively G → A or C → T conversions (Figure 3A, bottom). Thus, the elementary explanation of the observed asymmetric stretches of G → A or C → T transitions is fixation of random mutations by DNA polymerase due to erroneous recognition of $O^6$ alkyl guanine and subsequent segregation of daughter strands between individual cells.

2. Given this simple explanation of the observed asymmetric stretches of G → A versus C → T transitions, what is the cause of the observed switches between the alternative stretches? As the most parsimonious mechanism, we considered sister-strand exchange during replication. It is well known that DNA lesions, most notably EMS-induced 3-methyl adenine, 1-methyl adenine and 3-methyl cytosine (20), cause collapse of replication forks, and their repair often involves sister-strand exchange (6,21). A hybrid DNA molecule resulting from this process will contain parts of both daughter strands, separated by the site of recombination, and thus will carry the molecular records of lesions that have occurred on both parental DNA strands. The resulting switches between the G → A and C → T stretches would constitute a different kind of mutation segregation event that we observe in our system.

To test if homologous recombination is involved in this phenomenon, we next used a *recA* mutant of *E. coli* for EMS mutagenesis. The effect of EMS on viability of the *recA* mutant was significantly more pronounced compared to a wild type strain (10% survival versus 50-60% survival, respectively), as expected from the known contribution of homologous recombination to the repair of desintegrated replication forks induced by DNA lesions. Sequences of DNA extracted from three independent colonies confirm the hypothesis of recombination contributing to the phenomenon of switching. Strikingly, in one colony (Figure 3B, top), the number of switches was dramatically reduced (from an average of 8-10 to 2), whereas, in two other colonies, no switches were observed, and accordingly only one kind of transition (G → A or C → T) spanned the entire genome (Figure 3B, bottom left and right, correspondingly).

The above data strongly argue for involvement of homologous recombination in the generation of the switches. On the other hand, explained in this simple way, our observation of asymmetric stretches provides independent and transparent 'genetic' evidence for semiconservative DNA replication, complementary to the classic biophysical evidence (22).

3. 'Mixed' states are most economically explained by the presence of a mutagenic DNA lesion ($O^6$ alkyl guanine) in the cell-founder of the colony (Figure S2). That a lesion-containing strand could be expected among the cells at the moment of plating is consistent with the fact that the EMS-treated cells underwent around 2 divisions during this period (Figure S3), hence the original mutagenized DNA strands could not be significantly diluted by the newly synthesized DNA in the cell population at this time. Thus the observation of the 'mixed states' was not surprising, given our aim to limit the number of divisions to a minimum in order to avoid the loss of slightly deleterious mutations.

4. Concerning the unequal genome-wide distribution of mutations, the presence of impressive mutation bunching in *recA* colonies (indicated in Figure 3B, and for the statistical significance see Figure 2D) suggests that, although recombination could partially account for this phenomenon in the recombination-positive CC102 strain, it cannot be the only cause of bunching. On the other hand, this feature is also reminiscent of transcription bursts, a phenomenon observed over the last decade in studies of individual cells (23-27). The rate of mutation generation is determined by competition between repair and replication. Given that the molecular physics that underlies these processes is fundamentally the same as that of transcription, these processes could be subject to similar stochastic fluctuations. Therefore, one may expect that the balance between the rate of DNA synthesis and the efficiency of repair might also fluctuate significantly between different positions of genome in an individual cell, thus providing an explanation for the observed mutation bunching. Moreover, the scale of this bunching can surpass the size of the *E. coli* genome, for example when the number and/or size of bunches varies between cells. This should result in a non-normal distribution of mutation numbers between cells sequenced. This is in fact what we observe in our experiments (Figures 2B,D): the number of mutations in the CC102-derived colonies H1 (41) and H2 (34) is unusually far from the mean value 71.25, with the *p* values of non-normality ~1e$^{-4}$ and ~1e$^{-6}$, correspondingly, whereas the *recA*-derived genome R1 has 49 mutations, which deviates from the Poisson distribution with *p* value ~1e$^{-3}$.

To confirm the importance of the balance between repair and replication as the determining factor in mutation frequency, we kept mutagenized CC102 in a non-replicating state (PBS solution) overnight before plating. Three colonies were sequenced and showed dramatic deviation from the pattern previously observed. One colony showed a significant decrease (2 orders of magnitude) in the number of mutations (Figure 4A). Two other colonies had a strong preference for mutations near the replication terminus. This is consistent with the notion that the balance between repair and replication kinetics contributes to the genome-wide differences in the mutation density (Figure 4B). Given that we were treating exponentially growing cultures, most of the cells had the regions around the OriC already replicated. When put into the nutrient-lacking medium after EMS treatment, the cells can complete replication of the area close to the terminus, thus giving a chance for the lesions in this area to be converted to mutations before repair. In contrast, the area around the OriC has less chance to replicate again and there the lesions are more likely to be repaired before replication and mutation fixation.

The phenomenon defying an easy explanation is the consistently low mutation density in the regions of the genome posed at the OriC-Ter axis and its orthogonal axis (Figure 2C). We do not favor the explanation of this feature by a negative selection of lethal mutations, as the genome-wide profile of the differences between two wild strains of *E. coli* (MG1655 and O157) shows no preference for conserved positions in these regions (Figure S4). The explanation for the consistent deviation from a random distribution in these areas of genome most likely resides elsewhere, e.g., in some transient aspects of the cellular response to EMS treatment. First, the DNA in these regions might be differentially protected from EMS (e.g., due to DNA folding and intracellular location (28,29)). Second, the regions could differ in the efficiency of repair of the EMS-induced lesions. Further research will be required to clarify this issue.

## Discussion

The main novelty of the present study is in initiating genome-wide analysis of induced mutagenesis at the level of the individual cell. The traditional approaches, which typically rely on selection screens, are limited to analysis of few genetic loci, and often the results obtained with two different systems do not agree (1). The advantage of the modern high throughput technology is in allowing for the selection independent observation of mutations and recombination events, and of their distribution throughout the genome, by direct sequencing of several bacterial colonies and obtaining a mutation profile separately for individual genomes.

Non-randomness of mutation distribution is usually discussed in terms of mutational hot spots, first observed by Benzer on the T4 rII locus (2). Later, the mutation distributions were shown to have both hot-spot and random components (30). However, the use of selection limited these studies to comparisons between positions within a model gene. With our approach, we were able to observe several new types of non-uniform distribution patterns, now on a truly genomic scale: strong correlation between the "C to T" or "G to A" transitions between adjacent positions in the genome, as well as strong genome-scale variations in mutation density, either consistent for different genomes or else genome specific.

Whereas most of the observed distribution patterns can be explained by the known features of enzymatic processes (semiconservative DNA replication, homologous recombination, competition between replication and repair), the source of others remains to be elucidated. In this respect, mutation bunching represents a particular interest. Most of the bunching events in the wild type cells (CC102) could be due to homologous recombination; however, their observation in the *recA* mutants suggests an additional mechanism, reflecting the stochastic nature of genome dynamics at the single-cell level. As far as the repair efficiency is concerned, the fluctuations in the amounts of repair enzymes in individual cells (e.g., a single cell of *E.coli* could contain as few as 20 molecules of the methyltransferase MGMT, which removes alkyl groups from the modified guanine (31,32)) have been proposed previously as a source of transiently hypermutable phenotypes (33). We can now extend this notion to account for the mutation bunching, i.e., local variations in the numbers of repair proteins could be responsible for position-dependent variations in mutation density inside individual genomes. Consistently, we also observe non-normal distribution of mutation numbers between individual genomes sequenced. On the other hand, we cannot exclude another source of this cell-cell heterogeneity – the physiological state of *E. coli* growing in Luria-Bertani broth changes at an $OD_{600}$ of 0.3 (34), which could contribute to the differences between individual cells in their response to EMS treatment at this cell density.

Aside from DNA repair, could the local variations in the rate of DNA synthesis also contribute to the mutation bunching? A position-dependent fluctuation in DNA replication rate *in vivo* is practically impossible to observe on large cell populations. On the other hand, some crucial components of the replication machinery are present in limiting amounts in the bacterial cell – it has only 10-20 molecules of DNA-polymerase III, and no more than 4 copies of functional hexamers of DnaB helicase (35), crucial for replisome assembly and function. Thus, a possible role for fluctuating rates of DNA synthesis in mutation bunching cannot be discarded outright.

Several other phenomena should be mentioned. In addition to the consistent genome-scale variations in mutation density discussed in the Results section (Figure 2C), we are also intrigued by: 1) the peculiar mutation enrichment around OriC in the *recA* cells (Figure 3B), 2) occurrence, for every experimental group analyzed, of an 'outsider' colony (i.e., L1, H1 and R1) having a distribution pattern different from the rest of the group.

Concerning the first feature, the negative selection of mutation bunches in the context of exponentially growing cultures might be responsible for the observed mutation enrichment around the OriC in the *recA* cells. The location of lethal DNA damage might correlate with the mutagenic damage. In order for the cell to survive, at least one chromosome should survive the treatment. Since, in exponentially growing cultures, there are more DNA copies near OriC compared to the Ter region (Figure 1A, and S5 for a *recA* colony), cells that have a "bunch" near OriC will have a higher survival rate than those that have a bunch near Ter, thus increasing the chance to observe colonies with the OriC region enriched in mutations.

As far as the "outsiders" are concerned, establishing their noteworthiness would require accumulation of sequencing information that is beyond the scope of the present study. Their existence might also indicate the need, in future studies, of using better controlled physiological conditions; e.g., the cell cycle state could be an important factor. For example, the immediate ancestors of the L1 and H1 colonies might have been caught at the end of the replication cycle during the EMS treatment, leading to the repair process outperforming mutation fixation in the case of colony L1, and to preservation of a large portion of the damaged parental DNA strand in the case of colony H1. The half-replicated genome could be a factor in the case of the R1 colony, but additionally, *recA*-independent illegitimate recombination might be involved. Overall, the occurrence of the "outsiders" illustrates that, given the somewhat unique prehistory of every individual colony, the study of genome dynamics with our approach – in addition to the search for universal patterns and explanations – will also require an element of 'historic reconstruction', already familiar to biologists, although in a different – evolutionary – setting.

Our work can be put into the perspective of the methodological transition that modern biology is currently undergoing. Increasingly popular 'omics' approaches aim to measure all relevant characteristics of the object studied. But, concerning an individual cell, how far can the 'omics' methodology reach? Technological limitations restrict studies of individual cells to measurement of only a few of their properties (e.g., by flow cytometry and related methods). In this respect, DNA sequencing provides the largest amount of information about an individual cell (4.6 Mb in the *E. coli* genome), as compared to any of its other observable characteristics. Alas, DNA sequence barely varies among the descendants of a single bacterial cell, limiting the value of genome sequencing in the studies of dynamical processes in individual cells. Induced mutagenesis is a way to perturb and thus introduce more dynamics into the otherwise static object, thus allowing one to take full advantage of the high-throughput sequencing in studying processes other than transcription (such as replication, repair, recombination and their complicated interactions) at level of the individual cell. Our results demonstrate the value of new genomic technologies in addressing various aspects of intracellular dynamics at the single cell level, a topic that is becoming increasingly important in the light of recent advances in the studies of the stochastic nature of intracellular dynamics and cell individuality. They also pave the way for independent verification of the traditional assumptions that underlie studies of genetic variability.

## Materials and Methods

**Bacterial strains and mutagenesis**

The CC102 strain was obtained from Dr. M. Saparbaev (IGR, France). The *recA*- strain was DH5α ( *fhuA2 Δ(argF-lacZ)U169 phoA glnV44 Φ80 Δ(lacZ)M15 gyrA96 recA1 relA1 endA1 thi-1 hsdR17* ). Bacteria were grown in LB (Luria-Bertani) broth to OD 0.3, washed twice with PBS (Phosphate Buffered Saline, 137.93 mM NaCl, 2.67 mM KCl, 1.47 mM $KH_2PO_4$, 8.1 mM $Na_2HPO_4$, pH of 7.4, Invitrogen), then resuspended in PBS to the original density. To 2 ml of suspension 35 μl of EMS was added, and the cells were incubated for 45 min at $37^0$C. Cells were washed twice in PBS, resuspended in 2 ml of PBS, and 100 μl of suspension was added to 2 ml of LB. The cells were grown for 2 h at $37^0$C and plated on LB plates at different dilutions.

**Library preparation for Illumina sequencing**

10-500 ng of *E. coli* genomic DNA was processed according to the recommended Illumina protocol (see Supporting Materials and Methods). Two generations of Illumina technology were used in this work. GA1 was used to sequence the clones H1, H3, H4, H5, R1, R2, R3 and L1. GA2 was used to sequence H1, H2, H3, H6, L2 and L3. GA2 is more advanced technology in that is yields 3 times more coverage (average 80 reads covering a particular position instead of 25) and less noise. For the DNA sequenced with both GA1 and GA2, presented are the results of GA2 sequencing. The raw data were deposited in the Short Reads Archive at NCBI, (accession number: SRA008271.12).

**Mutations calling**

For the details of sequence alignment and the filtering criteria for the search of mismatches see Supporting Materials and Methods. Each mutation was attributed a "mixed state parameter" *r* – ratio of numbers of fragments carrying the mismatch versus total number of fragments covering this position. We found that to avoid significant contribution of false positive calls this parameters must be larger than 0.5. Notably false positive calls appear first in different from G → A and C → T mutation types as seen in H4, H5, L1 and R1 samples.

## Acknowledgements:

The authors thank Dr. M. Saparbaev and A. Ishchenko for discussion and helpful suggestions, Dr. A. Kuzminov and B. Hall for discussion, and Dr. L. Pritchard for critical reading of the manuscript. We thank the editor for the interpretation of the *recA* mutation distribution pattern and Dr. A. Danchin for the link with the Meselson and Stahl experiment. This work was supported by grants from "La Ligue Contre le Cancer" (9ADO1217/1B1-BIOCE) and the "Institut National du Cancer" (247343/1B1-BIOCE) to VO; and by Max Planck Society and the European Community's Seventh Framework Programme (FP7/2007–2013) under grant agreement # HEALTH-F4-2008-201418, entitled READNA (REvolutionary Approaches and Devices for Nucleic Acid analysis) to DP, VA and AS.

# Figures and legends

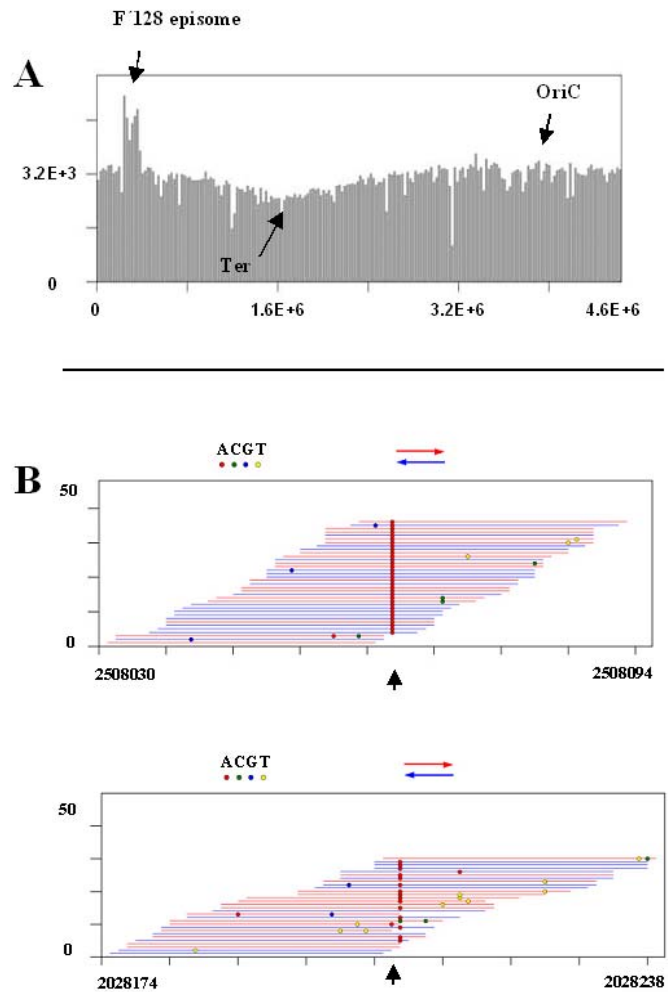

Figure 1

**Whole genome sequencing of individual colonies and mutation identification.**

**A.  An example of genome-wide coverage.** The sequence fragments obtained from the sequencing of DNA of an individual colony were aligned to the reference genome of the MG1655 strain. Abscissa, position along MG1655 genomic sequence. Ordinate, number of fragments per every 23198 bp (1/200 of the genome). The positions of the replication origin (OriC), terminus region (Ter) and F' 128 episome are indicated. The F' 128 episome is a plasmid that contains the chromosomal proB-lacZ region and independent replication origin, hence the discontinuity in the otherwise smooth gradient of genome coverage. The coverage (and correspondingly the amount of DNA) at Ter was about twofold smaller than that at OriC for all genomes sequenced. Thus there is one pair of replication forks per cell, in average exponentially growing cell.

**B.     An example of identification of individual mutations.** Reads of 23-32 bases were aligned with the reference sequence (parental CC102 strain). Differences with respect to the reference sequence are indicated by color. Noise is represented by colored positions that are unique to one sequence. Mutations, indicated by arrows, show a consistent difference throughout all reads (top), or through a significant part of the reads ('mixed positions', bottom).

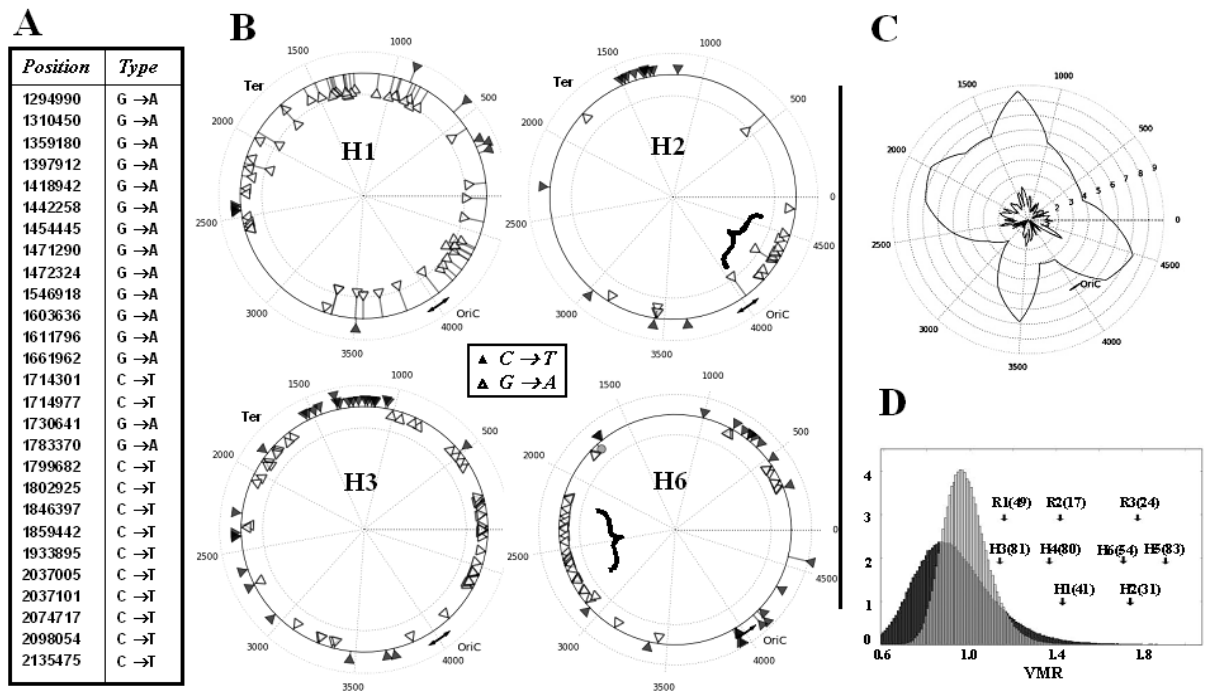

**Figure 2**

**Unexpected features of the mutation distributions.**

**A.   Example of G to A and C to T stretches.** Shown is the 1294990-2135475 region of the *E. coli* K12 MG1655 genome. The positions of mutations (left column) and their type (right column) are indicated. Presented are mutations observed after sequencing the genome of colony H1.

**B.   Genome-wide distribution of mutation type.** Results for four independent colonies are shown. The G → A, C → T and T → C transitions are indicated by open triangles, closed triangles and a circle, respectively. The 'pure' or 'mixed' state for every mutation is also indicated. The mutations with the 100% single nucleotide state are placed at the solid circle. The distance from the solid circle is proportional to the percent of wild type state detected; two dashed lines show 50% wild type state. Examples of mutation bunching (locations of increased mutation density that vary between different colonies) are indicated by "**{**".

**C.   Genome-wide distribution of mutation density.** Shown are the mutation densities obtained by averaging of data for the 6 genomes of the mutagenized CC102 strain sequenced. The genome was divided to 20 bins of 232 Kb size (coarse grained distribution, outer curve) or to 100 bins of 46 Kb size (fine grained, inner curve) and the mutation numbers in percent of total are plotted along the genome. Values closer to the center correspond to the regions of lowest mutation density.

**D.     Mutation bunching.** The term 'bunching' (and antibunching) is generally used to describe stochastic behavior which deviates from a random Poisson distribution, when successive events are not realized randomly but depend on neighboring events (36,37). Such behavior is widely observed in diverse settings from photon counting experiments to statistics of neuron firings. The departure from the normal distribution can be quantified by the variance to mean ratio (VMR, Fano factor). Here the statistics of distances between successive mutations in experimental samples is compared with simulated random mutations. The VMR distribution for 20 (black) and 80 (gray) random mutations in the *E. coli* genome was obtained by simulating half a million randomized mutagenesis acts. The distribution of distances between random mutations is binomial; thus its VMR is less than one. The experimental VMR values for different samples are shown by arrows, where H1-H6 corresponds to the mutagenized CC102 strain, and R1-R3 to the mutagenized *recA* strain. All our samples fell into the right tail of the distribution, some of them displaying VMR values highly unlikely for random mutations (P-value $\sim 1e^{-4}$).

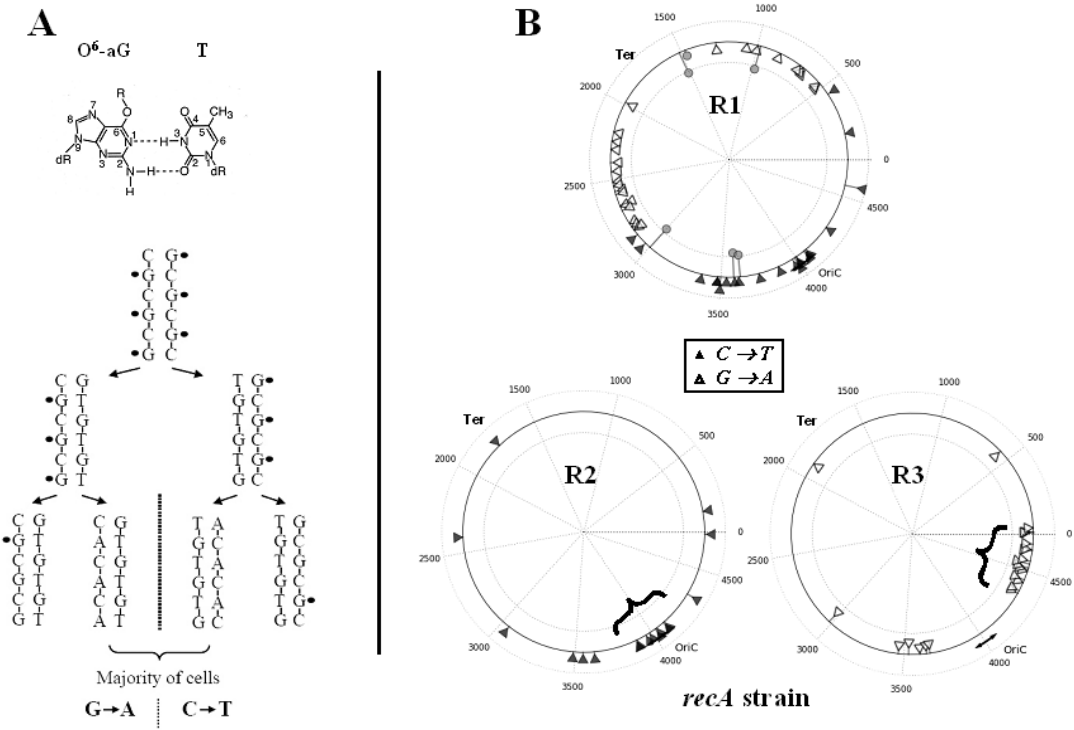

**Figure 3. Models and experimental verifications.**

**A.    Asymmetric stretches.**

**Top:** Scheme of the $O^6$ alkyl guanine ($O^6$-aG specifically mis-pairing with thymine (**T**), which should result in G:C → A:T replacement after a second round of replication.

**Bottom:** Model of generation of asymmetric stretches. For simplicity, the original sequence is depicted as consisting of G and C only, each G being alkylated by EMS treatment. After the first replication round, two daughter strands are generated, both carrying T paired with $O^6$-aG. After the second replication round, the DNA molecules with both newly synthesized strands carry exclusively either G → A (left) or C → T (right) replacements. Repair (for example, via removal of alkyl groups by methyltransferase MGMT) is also shown as conversion of "G*" back into "G".

**B.    Asymmetric stretches in the RecA background**. Genome-wide distribution of G → A (open triangle) and C → T (closed triangle) mutations for three colonies of *recA* mutant subjected to EMS treatment and processed as in Fig.1A. Locations of increased mutation density that vary between different colonies are indicated by "**{**". Mutations that are different from the G:C → A:T type are indicated by circle. Most of these positions have mixed state parameter *r* < 50%.

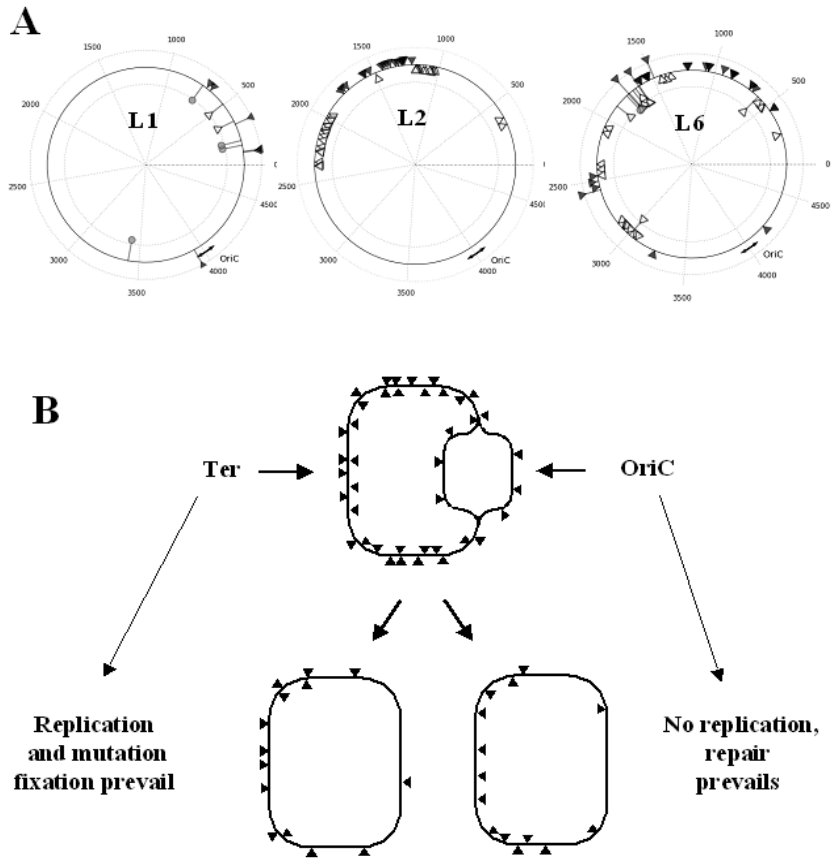

Figure 4

Role of competition between replication and repair in generation of mutations.

**A.    Starved cells.** Genome-wide distribution of mutations in CC102 strain cells kept in a non-replicating state (PBS) overnight after EMS treatment before plating. Three sequenced genomes are shown, with the mutation positions indicated as in Fig. 3B.

**B.    Model of competition between replication and repair.** Most of the cells in exponential culture have the regions around the OriC replicated. When put in the nutrient-lacking medium after EMS treatment, the cells can complete replication of the area close to the terminus, thus giving a chance for the $O^6$-aG in this area to be converted to mutations before repair. In contrast, the area around the OriC has less chance to replicate again and there the $O^6$-aG are more likely to be repaired before replication, thus avoiding mutation fixation.

| MG1655 versus CC102 | | CC102 WT versus CC102 Mut $r > 0.5$ | | CC102 WT versus CC102 Mut $r > 0.9$ | |
|---|---|---|---|---|---|
| A → C | 1 | A → C | 0 | A → C | 0 |
| A → G | 6 | A → G | 0 | A → G | 0 |
| A → T | 2 | A → T | 1 | A → T | 0 |
| C → A | 4 | C → A | 5 | C → A | 1 |
| C → G | 2 | C → G | 0 | C → G | 0 |
| C → T | 10 | C → T | 138 | C → T | 107 |
| G → A | 20 | G → A | 278 | G → A | 200 |
| G → C | 2 | G → C | 0 | G → C | 0 |
| G → T | 7 | G → T | 4 | G → T | 0 |
| T → A | 3 | T → A | 1 | T → A | 0 |
| T → C | 7 | T → C | 2 | T → C | 0 |
| T → G | 7 | T → G | 0 | T → G | 0 |

**Table S1**. **Summary of the single nucleotide differences.** The parental CC102 strain is compared with the reference strain K12 MG1655 (left column) and with the mutagenized CC102 colonies (center and right column). The type of mutation and the total numbers are indicated on the left and right sides of each column, respectively. The data for the two "mixed state parameter" values $r > 0.5$ and $r > 0.9$ (see Materials and Methods) are presented.

**Supplementary figures and legends**

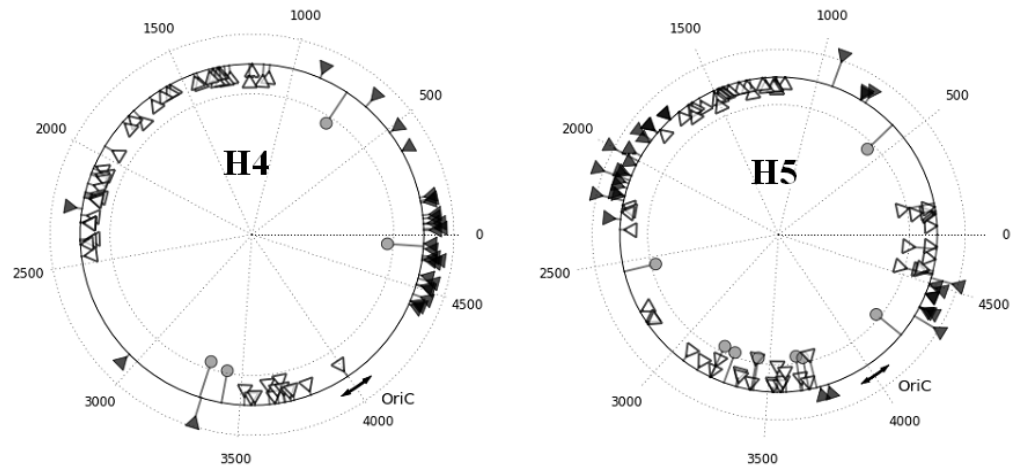

**Figure S1**

**Genome-wide distribution of mutation for the CC102 colonies H4 and H5.** The G → A, C → T and other transitions are indicated by open triangles, closed triangles and a circle, respectively. The 'pure' or 'mixed' state for every mutation is also indicated, as previously (Figure 2B).

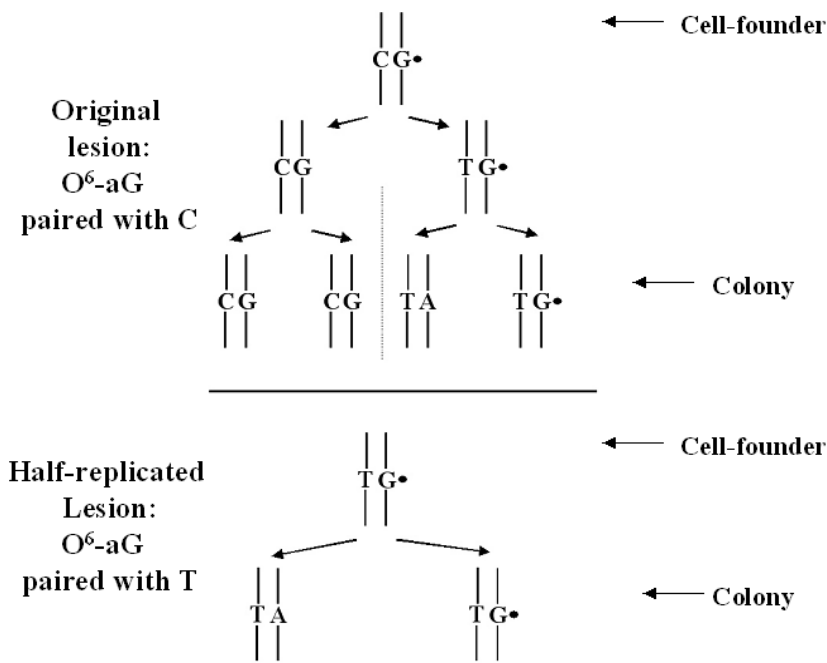

**Figure S2**

**Origin of 'mixed state'.**

We consider two simplified scenarios, which a). assume that every $O^6$-aG leads to T pairing during replication and b). do not take into account selection bias (both *in vivo* or during sequencing manipulations). **Top**. If no replication occurred between EMS treatment and plating, the founder cell of the colony will contain an $O^6$-aG:C pair. Assuming certain level of repair, more than 50% of the cells in the colony will contain the wild type sequence. **Bottom**. If only one replication round occurred between EMS treatment and plating, the founder cell of the colony will contain an $O^6$-aG:T pair. Assuming certain level of repair, the colony will contain between 50% to 100% of mutant sequences.

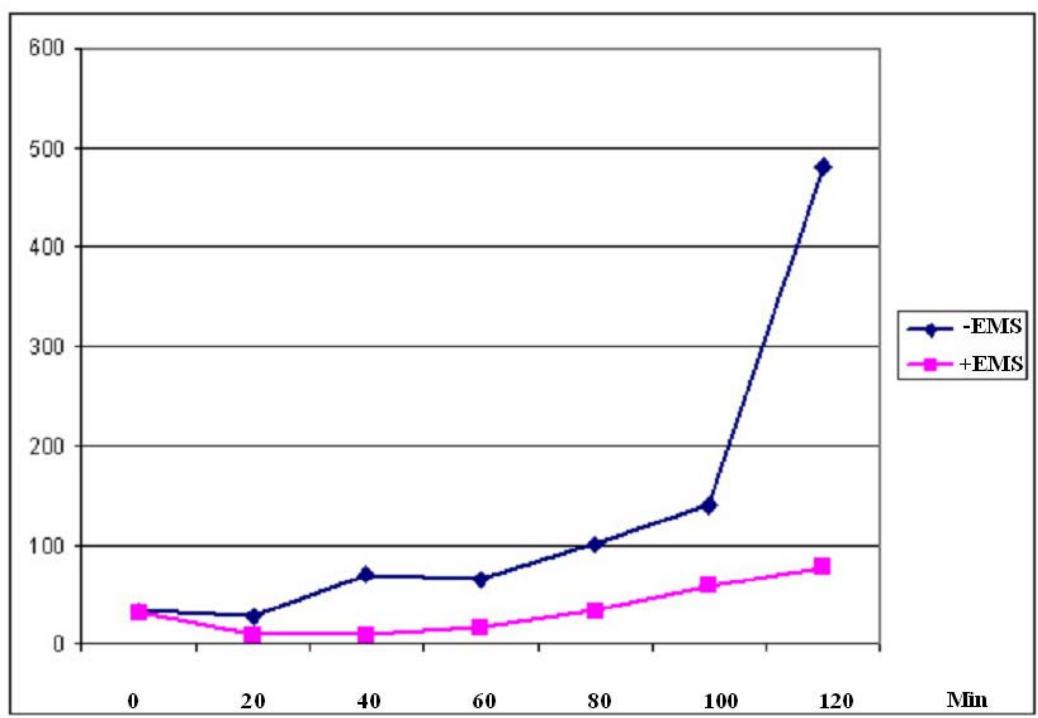

Figure S3

**Slow growth after EMS treatment.** 100 ul of cells were added to 2ml of LB medium and incubated at $37^0$C. Aliquots were taken out of the growing cultures at the times indicated on the abscissa and plated on LB agar at different dilutions. The ordinate indicates number of colonies per a plate for the $10^{-6}$ dilution.

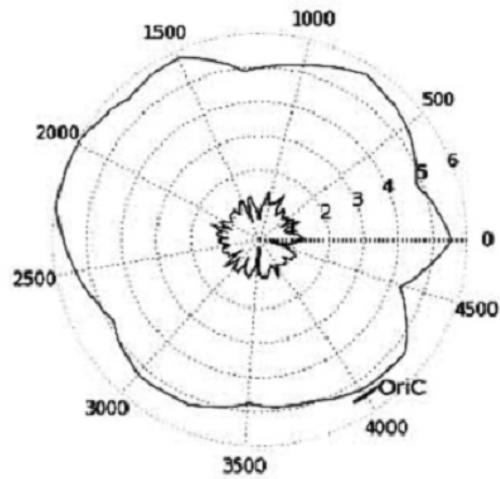

**Figure S4**

**No preference for conserved positions in regions of the genome posed at the axis orthogonal to the Ori-Ter axis.** Genome-wide distribution of the point differences between the *E. coli* strains MG1655 and O157. The genome was divided into 20 bins of 232 Kb size (coarse-grained distribution, outer curve) or to 100 bins of 46 Kb size (fine-grained, inner curve) and the mutation numbers in percent of total are plotted along the genome. Values closer to the center correspond to the regions of lowest mutation density.

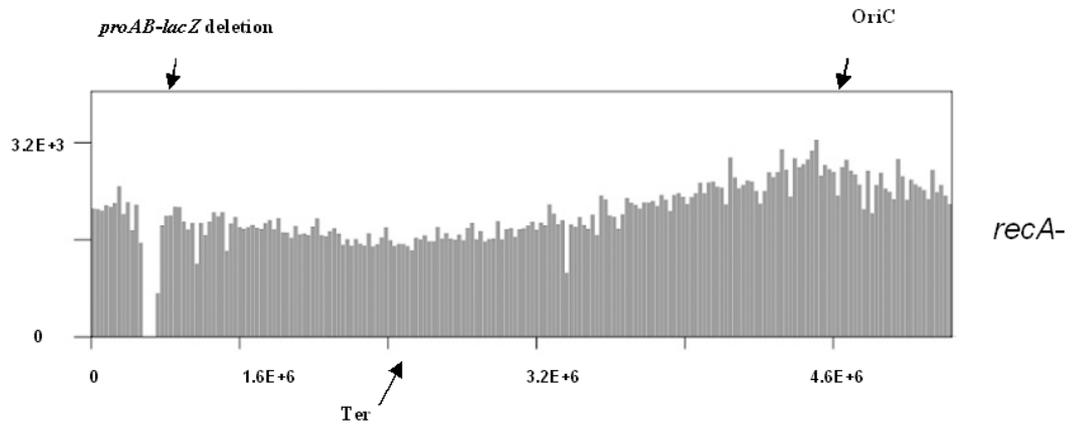

**Genome-wide coverage profile of a *recA* colony.** The sequence fragments obtained from the sequencing of DNA of the colony were aligned to the reference genome of the MG1655 strain. Abscissa, position along MG1655 genomic sequence. Ordinate, number of fragments per every 23198 bp (1/200 of the genome). The positions of the replication origin (OriC), terminus region (Ter) and the *proB-lacZ* deletion are indicated.